\documentstyle[12pt,equation,epsfig]{article}

\newcommand{\be}{\begin{equation}}
\newcommand{\ee}{\end{equation}}
\newcommand{\bea}{\begin{eqnarray}}
\newcommand{\eea}{\end{eqnarray}}
\newcommand{\nono}{\nonumber \\}
\newcommand{\la}{\label}
\newcommand{\ci}{\cite}
\newcommand{\bi}{\bibitem}
\newcommand{\half}{\frac{1}{2}}
\newcommand{\fourth}{\frac{1}{4}}

\newcommand\umu{^\mu}
\newcommand\lmu{_\mu}
\newcommand\p{\partial}
\newcommand\s{\sigma}
\newcommand\om{\omega}
\newcommand\f{\phi}

\newcommand\fover{\left({\f\over\f_0}\right)}

\begin{document}
\rightline{TAUP 2294-95}
\rightline{nucl-th/9510011}
\rightline{October 1995}
\vspace {1.5 true pc}
\centerline{{\large\bf Hot nuclear matter with dilatons}}
\vskip\baselineskip
\centerline{G. K\"albermann$^{\rm a}$, J. M.  Eisenberg$^{\rm b,c}$, 
and B.  Svetitsky$^{\rm b}$}
\vskip\baselineskip
\centerline{$^{\rm a}${\it Rothberg School for Overseas Students} }
\centerline{\it and Racah Institute of Physics}
\centerline{\it Hebrew University, 91904 Jerusalem, Israel}
\vskip\baselineskip
\centerline{$^{\rm b}${\it School of Physics and Astronomy}}
\centerline{\it Raymond and Beverly Sackler Faculty of Exact Sciences}
\centerline{\it Tel Aviv University, 69978 Tel Aviv,
Israel\/\footnote{Permanent address for J.M. Eisenberg}}
\vskip\baselineskip
\centerline{$^{\rm c}${\it GSI Darmstadt, Postfach 11 05 52}}
\centerline{\it D-64220 Darmstadt, Germany}

\begin{abstract}
\baselineskip 1.5 pc
We study hot nuclear matter in a model based on nucleon interactions
deriving from the exchange of scalar and vector mesons.  The main new
feature of our work is the treatment of the scale
breaking of quantum chromodynamics through the introduction of
a dilaton field.  Although
the dilaton effects are quite small quantitatively, they affect the
high-temperature phase transition appreciably.
We find that inclusion of the dilaton leads to a metastable high-density state
at zero pressure, similar to that found by Glendenning who considered instead
the admixture of higher baryon resonances.
\end{abstract}
\vfil
\newpage
\baselineskip 1.5 pc

\section{Background of the problem}

Following upon the early efforts \ci{MG,Wa1} to treat nuclei using 
relativistic mean-field theory with the exchange of scalar and 
vector mesons, a great deal of work was done
[3--8] to study the consequences of this approach for hot nuclear matter
and for neutron stars and supernovae \ci{ST,SKT}.  Within the last few
years it has been stressed \ci{RK,MBR} that since these approaches
to the many-nucleon strong-interaction problem must ultimately base
themselves on quantum chromodynamics (QCD) they should incorporate such
symmetries of QCD as chiral and scale invariance and the breaking of
these symmetries there at the quantum level.  Both references \ci{RK,MBR} 
treat
these issues at zero temperature, and it is our purpose here to explore
the consequences of incorporating broken scale invariance, or the QCD 
trace anomaly, at nonzero temperatures.
We are interested in particular in temperatures
approaching those of the deconfinement phase transition, namely, the range
of about 150 to 200 MeV.  We employ a scalar field, the
dilaton, to represent the effects of glueballs and to mimic the QCD 
trace anomaly, following a 
method originally introduced by Schechter \ci{Schech}.
We do not here consider the link between chiral features and the dilaton
considered by Mishustin, Bondorf, and Rho \ci{MBR}: Its main effect
is to displace the minimum of the dilaton field from its value
when chiral symmetry is respected and the location of that minimum is 
irrelevant in nuclear matter.

The phase transition found with the present model is obviously not the
confinement phase transition, since the degrees of freedom are entirely
hadronic.
Neither is it the phase transition of chiral symmetry restoration,
since our model has no chiral symmetry.
There may indeed be three separate phase transitions in the
150--200~MeV
temperature range, where the third is the transition studied here:
a sharp reduction in the nucleon's effective mass, accomplished by the
self-coupling of the concomitant high density of nucleon-antinucleon
pairs.
A more extensive model, with the missing degrees of freedom, will show
whether two or more of these transitions might coalesce into, say, a
single quark--hadron transition.\footnote{For
a discussion of the relationship between confinement and chiral
symmetry phase transitions, see \cite{Sv}.}

Moreover, since the chiral transition is the only one associated with
an (almost) exact symmetry, it is the only one which assures us that we
{\em must} encounter an actual singularity as we raise the temperature and/or
the density.
As we shall see, the phase transition studied in this paper occurs (as
a function of temperature) only for a limited range of densities.
Elsewhere the high- and low-temperature regimes are continuously
connected.

Our motivation for studying dilaton effects in hot nuclear matter is
to explore the consequences of the fundamental scaling features
of QCD within a mean-field model that has had much success in its nuclear
applications \ci{SW}.  The criticism has been voiced \ci{Birse} that dilaton
effects are easily overestimated in nuclear applications if too small a
mass is attributed to the glueball in place of values for it in the
expected range of 1.5 to 2 GeV.  We anticipate that this issue
is irrelevant here, because our use of the dilaton in nuclear matter
does not require the fixing of a value for the glueball mass.
Moreover,
quite small changes in energy balance within the regions of phase
transition can lead to different qualitative behavior.
We find signs of that here.

As this work was being finished, we learned of a parallel calculation
being carried out by Papazoglou \ci{MPS} which reaches 
similar general conclusions about the role of the dilaton in hot nuclear
matter (and also considers extensions to nonzero strangeness).  
That work considers a particular choice of dynamics for
the inclusion of chiral invariance and the nucleon mass is
generated by spontaneous symmetry breaking.
It is interesting to compare those results with ours since
nuclear thermodynamics are very sensitive to any change in the nucleon
effective mass.

\section{Formalism}

Formal features for the handling of hot nuclear matter in mean-field
theory have been presented frequently
[3--5, 7--12]
and we here sketch them only very briefly in order to
establish notation and procedures.  Our lagrangian is \ci{MBR,Schech}
\bea \la{L} 
{\cal L} & = & \bar\psi\left[i\gamma\umu\left(\p\lmu+ig_\om\om\lmu\right)
- \left(\fover M+g_\s\s\right)\right]\psi  
+ \half\p\umu\s\p\lmu\s - V[\s,\,\f]  \nono
&&-\fourth\big(\p\umu\om^\nu-\p^\nu\om\umu\big)
\big(\p\lmu\om_\nu-\p_\nu\om\lmu\big)  
+ \half m_\om^2 \fover^2 \om\umu\om\lmu  \nono
&&+\half \p\umu\f\p\lmu\f 
- U[\f]\ ,
\eea
where $\psi$ is the nucleon Dirac field
and $\om\lmu$ is the field for the
vector meson.
We use
the usual fourth-order polynomial form for the potential of the
scalar field $\s$ in the presence of the dilaton $\f,$
\be \la{V}
V[\s,\,\f] = \half m_\s^2 \fover^2\s^2 + {1\over 3} g_3 \fover \s^3
+ \fourth g_4 \s^4.
\ee
The dilaton potential
\be \la{U}
U[\f] = B\left(e\frac{\f^4}{\Lambda^4}
\log{\f^4\over \Lambda^4} + 1 \right)
\ee
yields the usual
classical minimum for the field $\f_0 = \Lambda/e^{1/4},$ where
$e \approx 2.718.$
The coupling strengths for nucleon-meson couplings are $g_\s$ and $g_\om$
for the scalar and vector fields respectively, and the masses of these
fields are $m_\s$ and $m_\om.$  We also require as input parameters 
here the coefficients $g_3$ and $g_4$ in the scalar potential.
The powers of $\f$ appearing in eqs.~(\ref{L}) and 
(\ref{V}) are
determined, as usual, by the dimensionality of the term in question.

The mean-field equations in infinite nuclear matter are given by
\be \la{scalar}
g_\s \rho_s + m_\s^2\fover^2 \s + g_3\fover \s^2 + g_4 \s^3 = 0
\ee
for the scalar field,
\be \la{vector}
g_\om \rho_v - m_\om^2\fover^2 \om = 0
\ee
for the vector field with only the time component $\om$ present, and
\be \la{nucleon}
M\rho_s + {1\over 3}g_3\s^3 + 4B\fover^3\left(1 +
\log{\f^4\over\Lambda^4}\right) = m_\om^2\fover\om^2 - m_\s^2\fover\s^2
\ee
for the nucleon.
Here the nuclear scalar and vector densities for nonzero temperatures are
\be \la{rho_s}
\rho_s = 4 \int{d^3 p\over (2\pi)^3}{M^*\over\sqrt{p^2+M^{*2}}}
\left(n_p^+ + n_p^-\right)
\ee
and
\be \la{rho_v}
\rho_v = 4 \int{d^3 p\over (2\pi)^3}\left(n_p^+ - n_p^-\right)\ .
\ee
The factor 4 appearing in eqs.~(\ref{rho_s}) and (\ref{rho_v}) is for
nucleon degeneracy.  The fermion and antifermion distributions are
\be \la{n}
n_p^\pm = {1\over 1 + \exp\left[\left(\epsilon_\pm(p)\mp\mu\right)/T\right]}\ .
\ee
Here the nucleon and antinucleon energies are given by
\be \la{eps}
\epsilon_\pm = \sqrt{p^2 + M^{*2}} \pm g_\om\,\om
\ee
and the nucleon effective mass is
\be \la{M_eff}
M^* = \fover M + g_\s\,\s.
\ee
These equations are now solved for given values of the temperature.
Once this is accomplished, the system energy density is
\bea \la{energy}
{\cal E} & = & 4\int{d^3 p\over (2\pi)^3}\sqrt{p^2+M^{*2}}
\left(n_p^+ + n_p^-\right) 
+ V[\s,\f] + U[\f] \nono
&&+\half \Bigg(m_\om\fover\Bigg)^{-2} g_\om^2\rho_v^2\ ,
\eea
and the pressure is
\bea \la{pressure}
P & = & {1\over 3}\cdot 4\int{d^3 p\over (2\pi)^3}
{p^2\over\sqrt{p^2+M^{*2}}} \left(n_p^+ + n_p^-\right) - V[\s,\f] - U[\f]
\nono 
&&+\half\Bigg(m_\om\fover\Bigg)^{-2} g_\om^2\rho_v^2\ .
\eea

Thermodynamic properties of hot nuclear matter are most easily discussed
in terms of the three intensive variables that characterize the mixed
phases near a transition, namely, the pressure $P$, temperature $T$, and 
chemical potential $\mu$.  This immediately recommends the use of the
grand potential \ci{TW,Ku}
\be \la{grand}
\om(\mu,\,T) = {1\over V}\Omega(\mu,\,T) = -P\ .
\ee
Near a phase transition we then have thermodynamic equilibrium for that
branch having highest pressure (minimal grand potential) at given
chemical potential and temperature.
Thus we show our results as functions of chemical potential
rather than baryon density so that transition points may easily be read
off the pressure curves and applied to the other thermodynamic variables.

At the temperatures of interest, a pion gas coexists with the nucleon
gas we are studying.
Its mean field is zero, and it affects the nucleon gas through
scattering.
Its treatment is beyond the scope of our mean-field approach.

\section{Results and discussion}

We fix parameters, basing ourselves on previous work \ci{ST},
such that at nuclear saturation the density is $\rho_0=0.153$~fm$^{-3}$,
the binding energy per nucleon is $E_B=-16.5$~MeV, and the
compressibility is $K = 212$~MeV when the dilaton is not present and 
$K = 234$~MeV when it is.  The nucleon's effective mass is $M^*=0.57\,M$.
The parameters of eqs.~(\ref{L}) and (\ref{V})
are then $g_\s = 10.1$, $g_\om = 13.3$, $m_\s = 492$~MeV, and
$m_\om = 795$~MeV for the scalar and vector mesons, and $g_3 =
-12.2$~fm$^{-1}$ and $g_4 = -36.3$ for the scalar potential, for
the case where the dilaton is absent.
When it is present the scalar meson
coupling becomes $g_\s = 9.6$ and the ``bag'' constant for the dilaton
field is taken to be $B = (200~{\rm MeV})^4,$ as in \ci{RK}.  \footnote{The
relative ease with which successful parameters are found in this model comes
 about because we have not constrained it to fit properties
of finite nuclei \cite{FS,HRE}.
We are satisfied with studying dilaton effects on a model which describes
well zero-temperature nuclear matter, and thus we have not explored
the parameter space more fully.} For
nuclear matter the value of the dilaton mean field $\f_0,$ or 
alternatively of the glueball mass, is not required since we express all
quantities in the equations of motion in terms of the ratio $\f/\f_0.$
We note that the scalar potential $V$ of eq.~(\ref{V}) used here suffers
from the problem of ``bifurcation,'' which is well known in this area
\ci{BS,WMSG}:
In order to obtain reasonable values for the
nuclear compressibility $K$ we are obliged to use negative, and therefore
destabilizing, coefficients $g_3$ and $g_4.$
One must keep in mind that our Lagrangian is after all that of
an effective field theory.

Figure~1 shows the well-known \ci{Glen,Wa2} gas--liquid phase 
transition that occurs in these mean-field models at low temperatures.
Not surprisingly, there is hardly a discernible difference between the
cases with and without dilaton for these low values of $T$.

We now consider the higher-temperature regime
where there appear larger
quantitative differences between the two cases. 
In the theory {\it with} dilatons, there is a first-order phase
boundary in the $T$--$\mu$ plane extending from the $T$ axis
at $T\simeq185$~MeV to a critical point at $(T,\mu)\simeq(170~{\rm MeV},
\,565~{\rm MeV})$.
In the theory {\it without} dilatons, the intercept is at\footnote{Due to
numerical difficulties we have not really verified that the phase boundary
intersects the $T$ axis in the theory without dilatons.}
$T\simeq189$~MeV, with a critical endpoint at $(T,\mu)\simeq(185~{\rm 
MeV},\,300~{\rm MeV})$.
Figures~2 and 3 show isotherms of pressure
and energy density as functions of chemical potential.
The phase transition is clearly visible as a crossing of
pressure curves and as a discontinuity in the energy density.
To characterize the phases we display the effective nucleon mass
in Fig.~4.
The low $(T,\mu)$ regime is a {\it normal} phase, by which we mean a
state with an effective nucleon mass
close to that of zero-temperature, ordinary nuclear matter.
The phase at large $T$ or large $\mu$ is an {\it abnormal} phase in that the 
nucleon's effective
mass is much smaller.
For $T=100$ and 150~MeV, one sees a smooth crossover between the two
regimes as $\mu$ is varied;
a phase transition appears in the isotherms when one crosses the phase
boundary just described.
For sufficiently large $T$ the system is abnormal for all $\mu$.

Since there are small changes in the parameters between the models with
and without dilatons one
should not take the differences in temperature
very seriously.  On the other hand, 
the abnormal states above the transition region show a rather
smaller effective mass (Fig.~4) in the presence of the dilaton than 
they exhibit
in its absence.  This is because when the dilaton is involved it provides
an additional mechanism to that of the scalar field for reducing the
nucleon mass above the transition region, as may be seen in 
eq.~(\ref{M_eff}).  The low effective mass that is found at and above the
transition region is the characteristic feature of the abnormal branch.
The occurrence of this abnormal feature is a consequence here \ci{Glen}
of a feedback loop whereby at high temperatures many nucleon-antinucleon
pairs may be produced, which in turn enhances the scalar field [see
eqs.~(\ref{scalar}) and (\ref{rho_s})], thus further lowering the effective
mass.  In the presence of the dilaton, the
$\f$ field also participates in this loop through eq.~(\ref{nucleon}).
As is well known \ci{LW}, there is an alternative mechanism which
produces an
abnormal state at {\it low} temperatures purely through the increase in baryon
density.

The most important consequence of the dilaton is in the metastable
extension of the abnormal phase to low values of $\mu$.
As seen in Fig.~2, both with and without the dilaton the metastable
state persists all the way down to $\mu=0$ (whence it continues into
the $\mu<0$ region since $\mu\to-\mu$ is a symmetry of the theory).
In the model {\it with} the dilaton, however, the metastable state crosses
$P=0$ which implies that a nugget of this phase is mechanically stable.
This feature was discovered by Glendenning \cite{Glen} in a model
{\it without} dilatons which included the contributions of a ladder
of baryonic resonances to the thermodynamics.
We can only speculate that inclusion of {\it both} the dilaton and
the baryonic resonances will serve to drive the zero-pressure state
towards higher $\mu$ and hence closer to the phase transition,
thus improving prospects for observability.
We note, however, that the metastable states in Fig.~2, for both
models, are separated from the true minimum of the grand potential at the
same $\mu$
(the stable phase) by a very low barrier.
This does not make us sanguine regarding the lifetime of the state.

We also present curves for the baryon density (Fig.~5)
and the dilaton field expectation value (Fig.~6).
The latter shows a slow decrease
as $\mu$ increases and then a jump to the
abnormal state in which the dilaton field drops by some 10\%.

In sum it appears that the incorporation of scale-invariance breaking in
a form suggested by QCD can have significant effects in the phase
structure of hot nuclear matter.  We hope that this observation may be of
particular use in the study of effects of a hot hadronic medium on baryon
behavior.  This has been explored, for example, using skyrmions \ci{Mish},
though thus far without consideration of dilaton effects.  Since such
effects have been found to be important for one- and two-skyrmion 
systems \ci{EK} it will be important to take the dilaton into account in
both its roles in eventual studies of baryons immersed in hot matter.
\vskip 2 true pc

It is a pleasure to acknowledge very useful conversations on the subject
matter of this paper with Walter Greiner, Igor Mishustin, Panajotis
Papazoglou, and Horst St\"ocker.  Part of this work was carried out while
JME was visiting the Institut der Theoretische Physik der
Universit\"at Frankfurt, and he wishes to express his gratitude for the
very kind hospitality he received there.
BS thanks the Weizmann Institute of Science for its continuing
hospitality.
This work was funded in part by the Israel Science Foundation, by the
U.S.-Israel Binational Science Foundation, and by the Ne'eman Chair in 
Theoretical Nuclear Physics at Tel Aviv University.

\vskip 2 true pc

\newpage
\parindent 0 pc
\parskip 0.5 pc
{\Large\bf Figure captions}

Figure 1.  Pressure $P$ versus chemical potential $\mu$ at low
temperatures $T = 8$ MeV (right) and 10 MeV (left) in the 
region of the gas--liquid phase transition: (a) no dilaton,
(b) with dilaton.
Dashed segments represent metastable states.

Figure 2.  Pressure $P$ versus chemical potential $\mu$ at the high-temperature
phase transition:
(a) without dilaton, at $T=188$~MeV; (b) with dilaton, at $T = 175$~MeV.

Figure 3.  Energy density ${\cal E}$ versus chemical potential $\mu$ for
values of temperature spanning the phase transition region:  (a)
No dilaton.
The isotherms represent, from bottom to top, $T = 100,$ 150, 186, 188, and
190~MeV.
(b) With dilaton.  Temperatures, from bottom to top, are $T = 100,$ 150, 172, 
175, and 200 MeV.

Figure 4.  Nucleon effective mass $M^*$ versus chemical potential $\mu$
for values of temperature spanning the phase transition region: (a) no dilaton,
(b) with dilaton.
The temperature values are the same as in Fig.~3, but in the opposite order:
top to bottom.

Figure 5.  Baryon density $\rho_v$  versus
$\mu$: (a) no dilaton, $T = 188$ MeV; (b) with dilaton, $T = 175$ MeV.

Figure 6.  Dilaton field $\f/\f_0$ for $T = 175$ MeV.

\newpage
\begin{figure}[t]\vspace{3cm}
\centerline{\psfig{figure=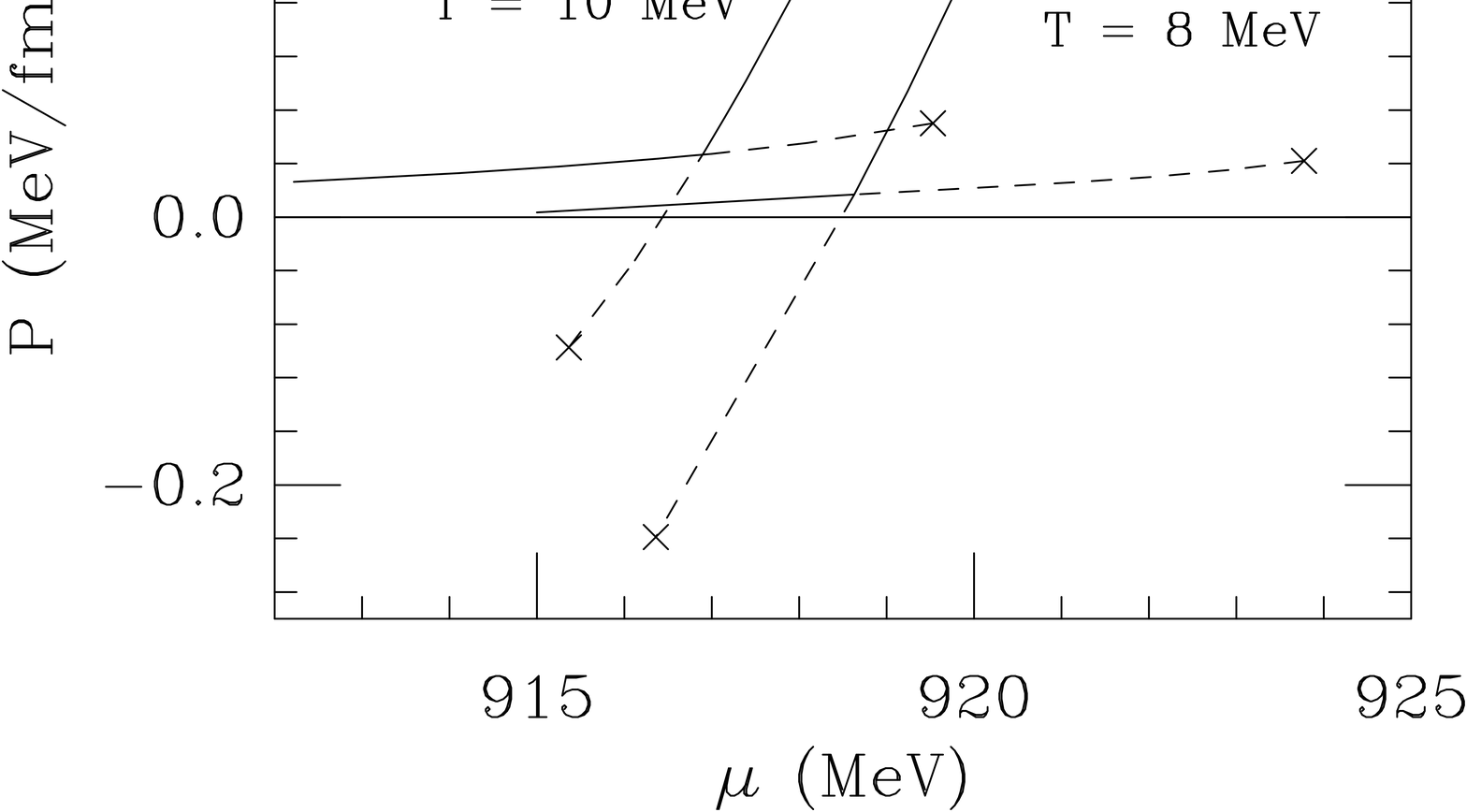,height=6cm}}
\centerline{Fig.~1(a)}
\end{figure}
\begin{figure}[b]\vspace{3cm}
\centerline{\psfig{figure=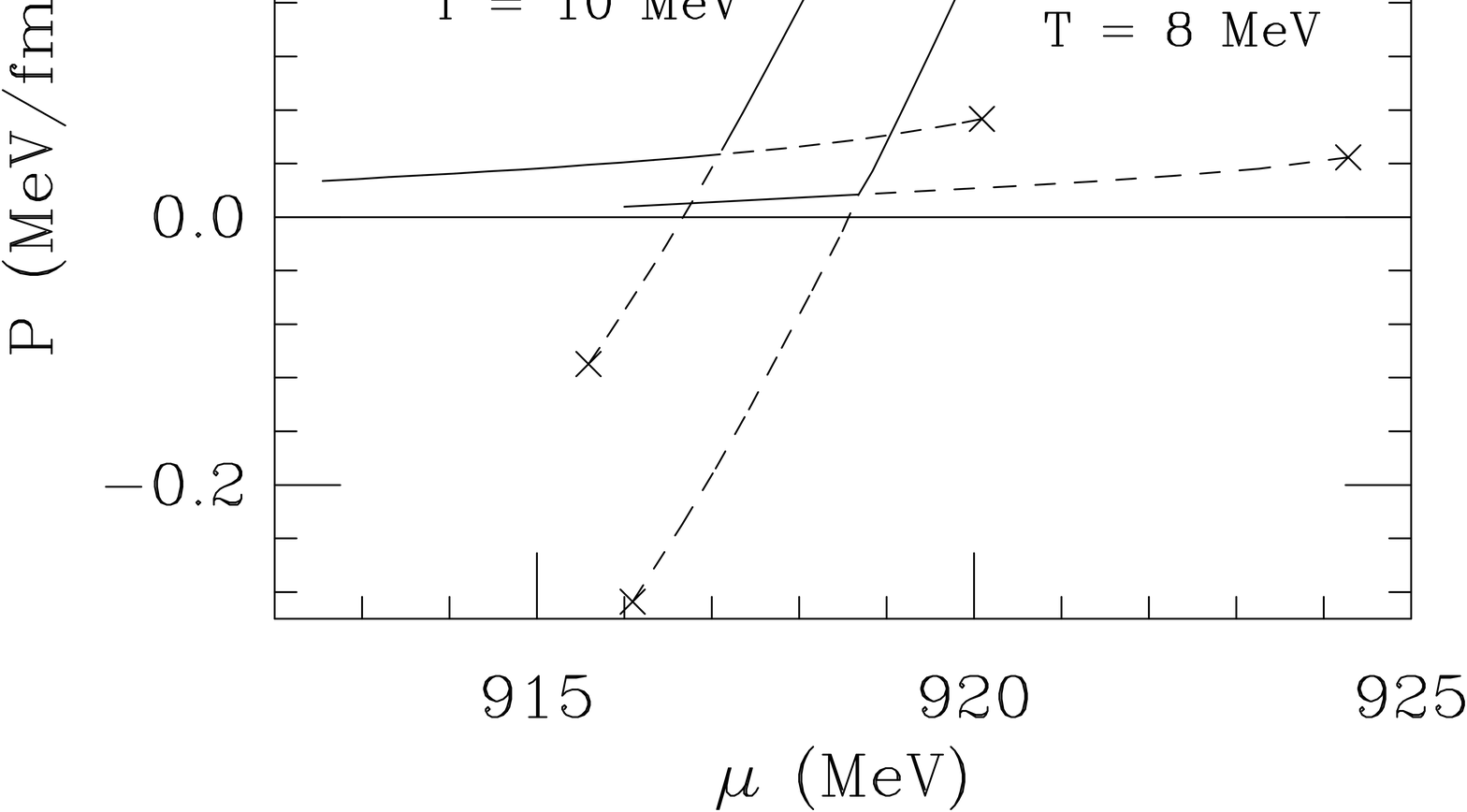,height=6cm}}
\centerline{Fig.~1(b)}
\end{figure}
\eject
\begin{figure}[h]\vspace{3cm}
\centerline{\psfig{figure=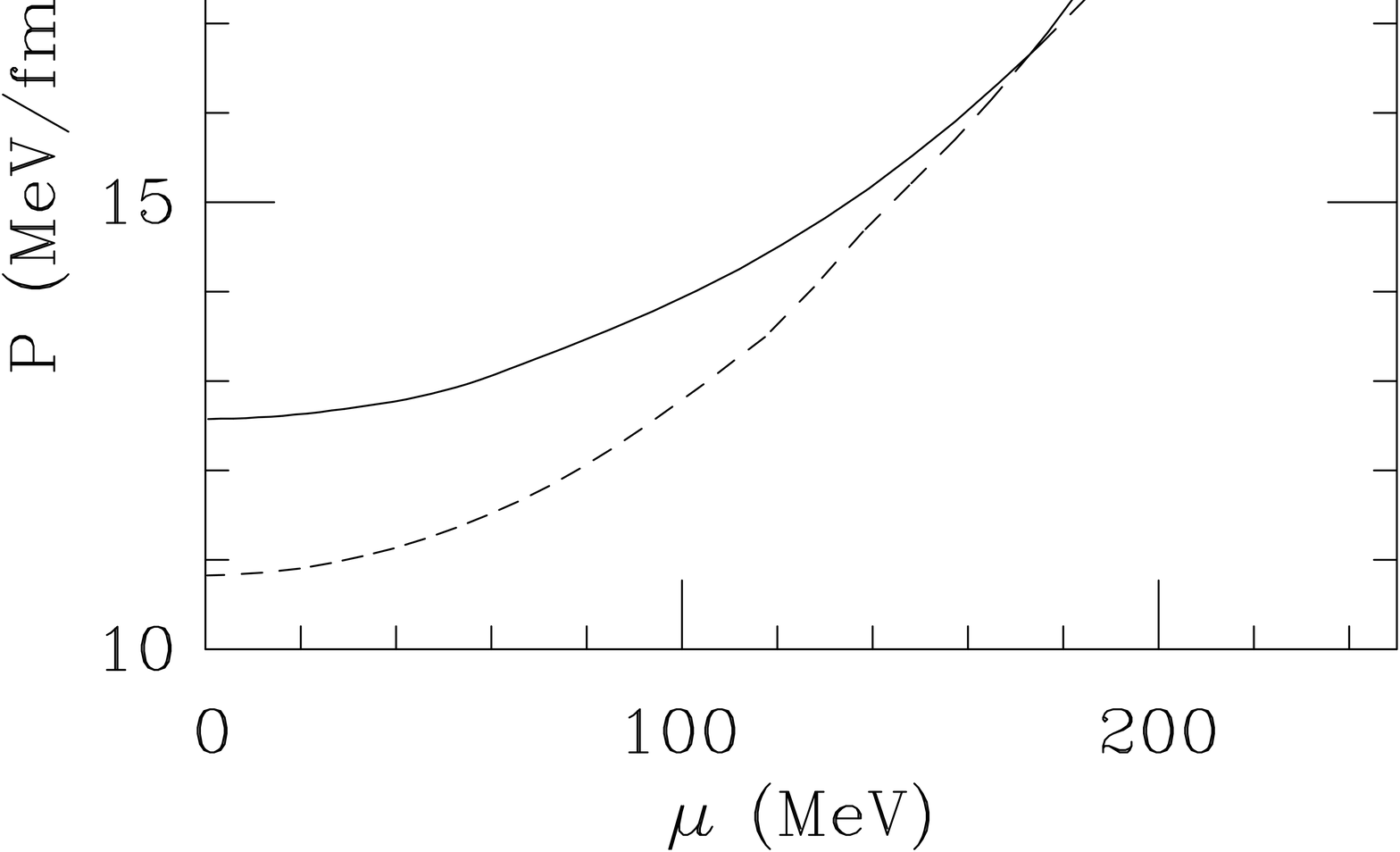,height=6cm}}
\centerline{Fig.~2(a)}
\end{figure}
\begin{figure}\vspace{3cm}
\centerline{\psfig{figure=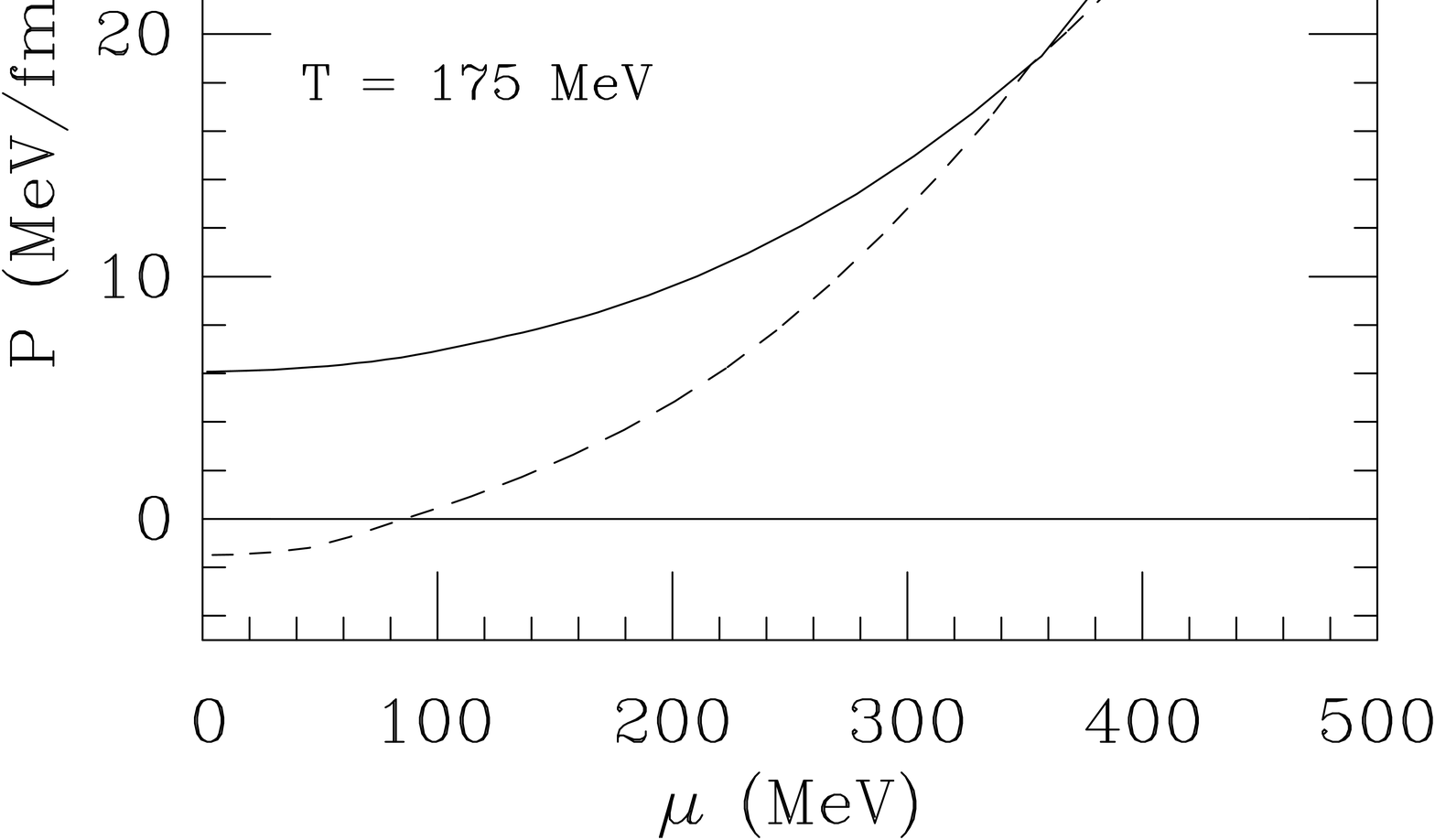,height=6cm}}
\centerline{Fig.~2(b)}
\end{figure}
\newpage
\begin{figure}[t]\vspace{3cm}
\centerline{\psfig{figure=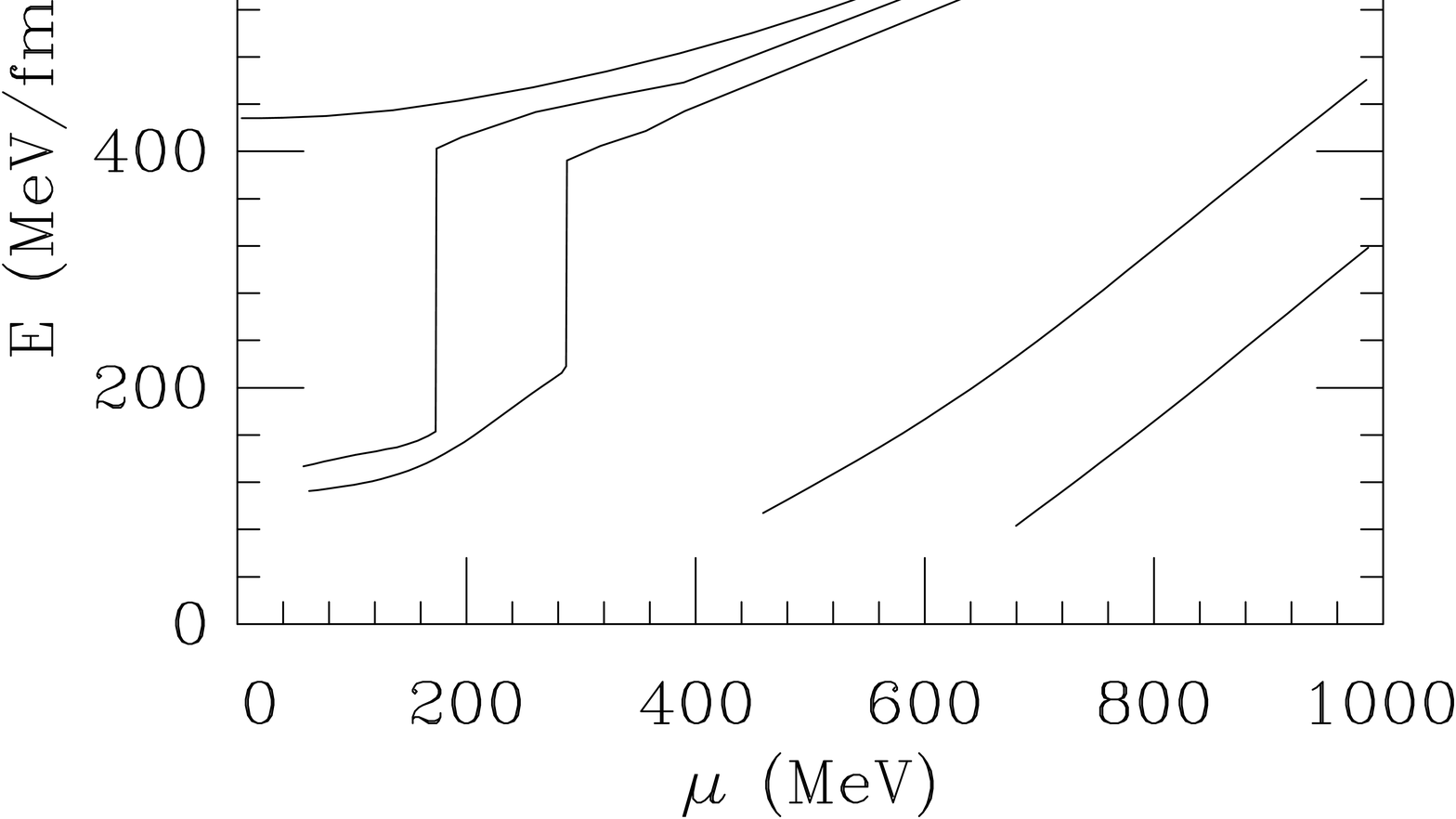,height=6cm}}
\centerline{Fig.~3(a)}
\end{figure}
\begin{figure}\vspace{3cm}
\centerline{\psfig{figure=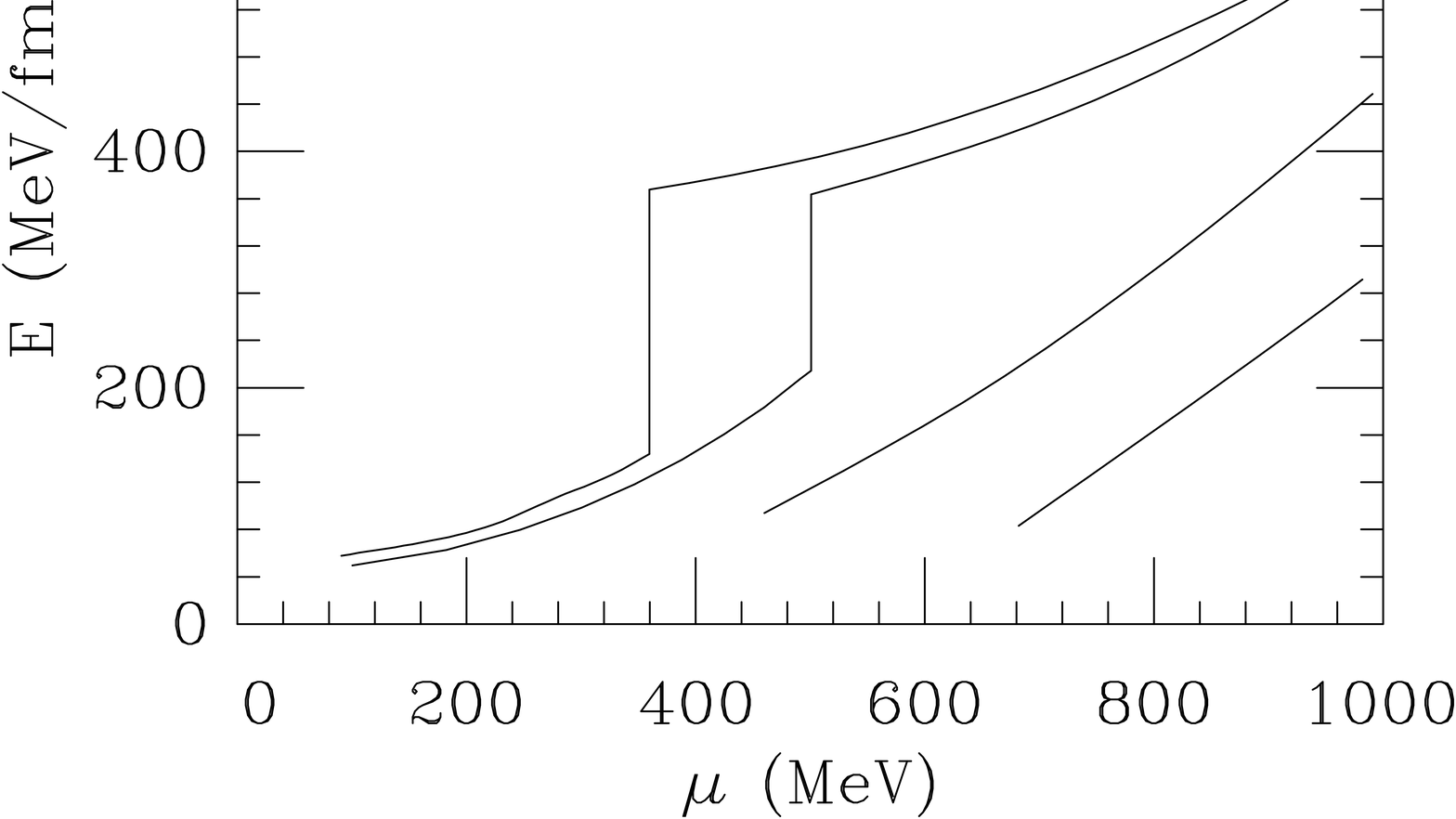,height=6cm}}
\centerline{Fig.~3(b)}
\end{figure}
\newpage
\begin{figure}[t]\vspace{3cm}
\centerline{\psfig{figure=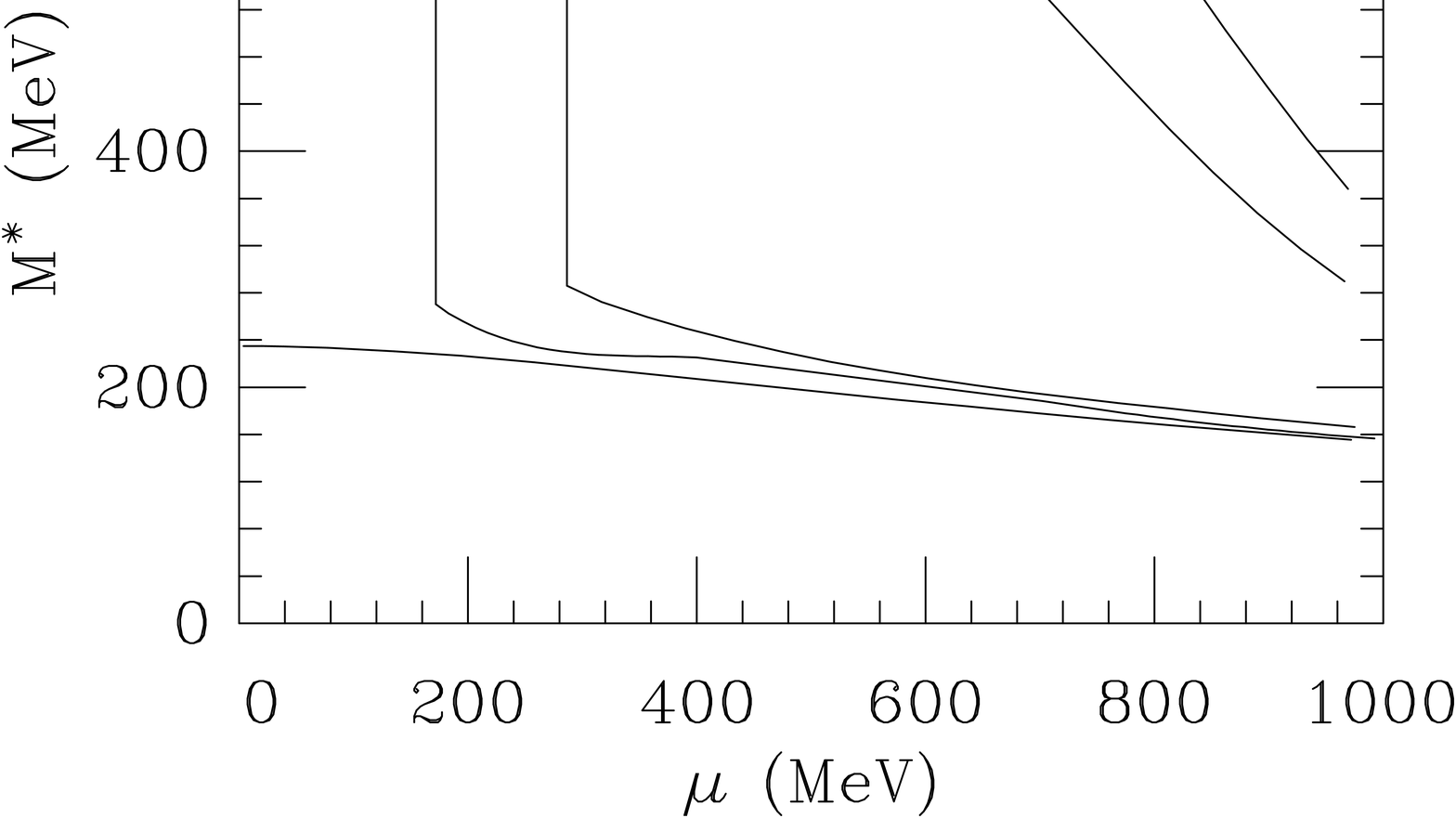,height=6cm}}
\centerline{Fig.~4(a)}
\end{figure}
\begin{figure}\vspace{3cm}
\centerline{\psfig{figure=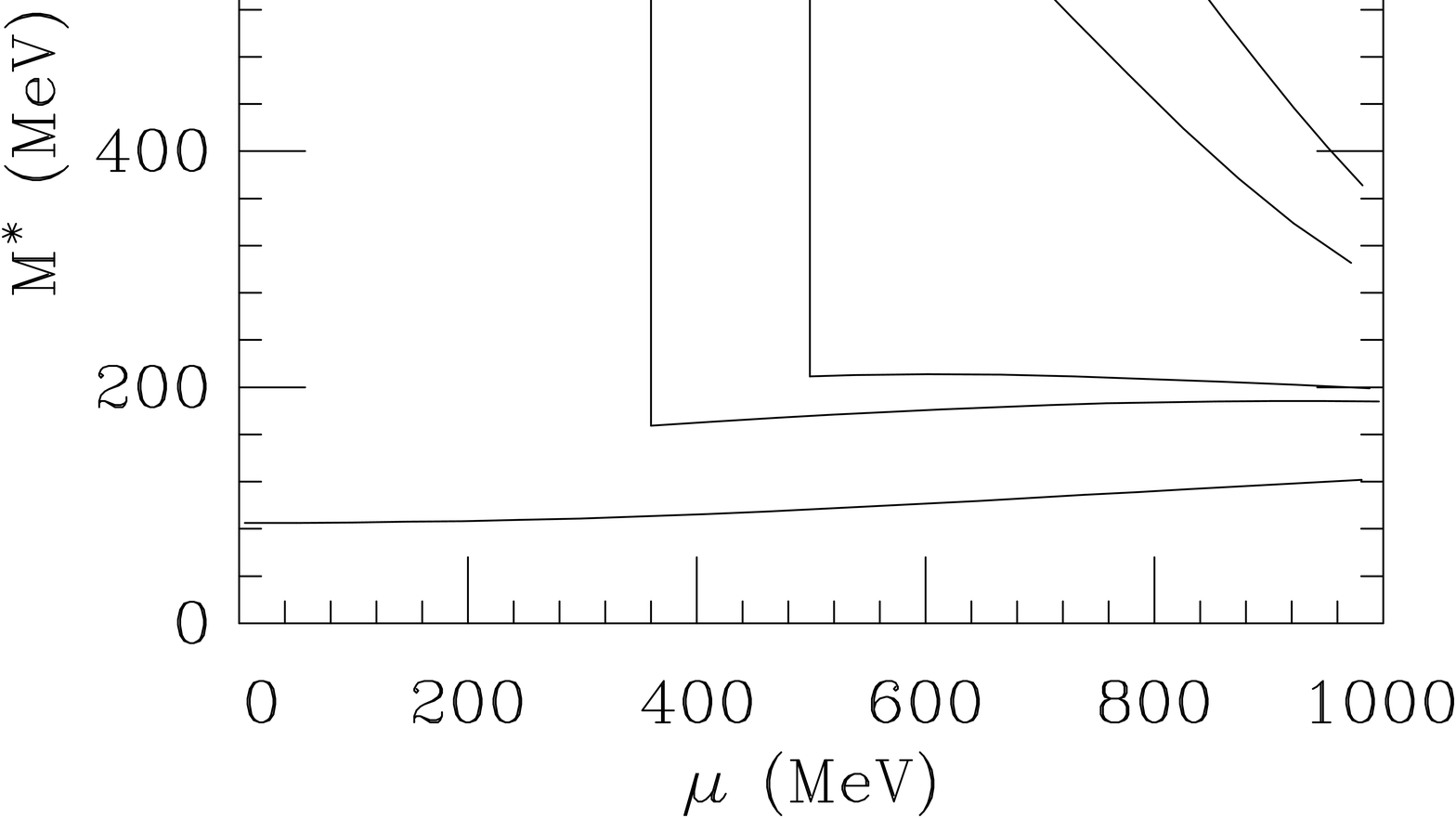,height=6cm}}
\centerline{Fig.~4(b)}
\end{figure}
\newpage
\begin{figure}[t]\vspace{3cm}
\centerline{\psfig{figure=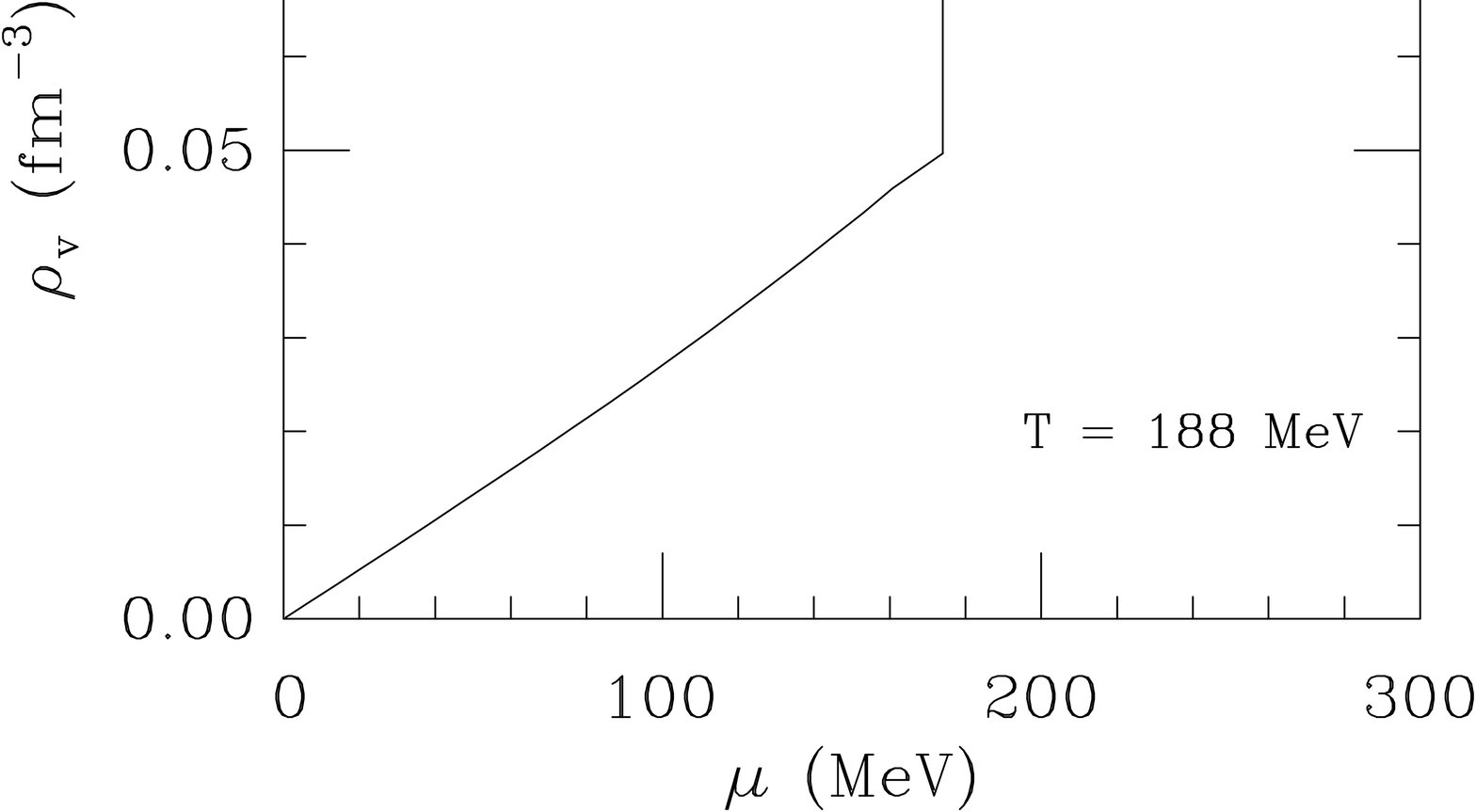,height=6cm}}
\centerline{Fig.~5(a)}
\end{figure}
\begin{figure}\vspace{3cm}
\centerline{\psfig{figure=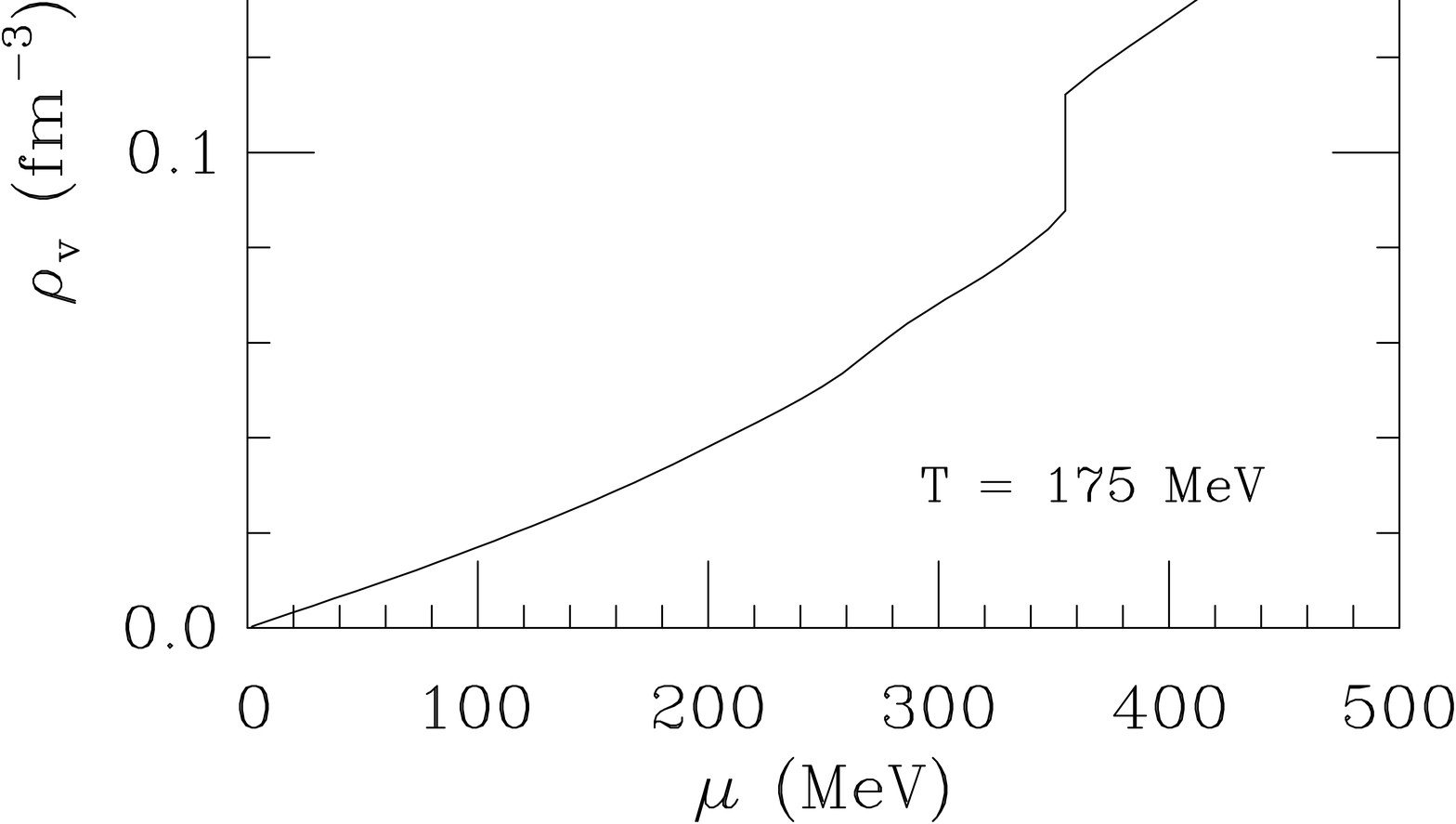,height=6cm}}
\centerline{Fig.~5(b)}
\end{figure}
\newpage
\begin{figure}[t]\vspace{3cm}
\centerline{\psfig{figure=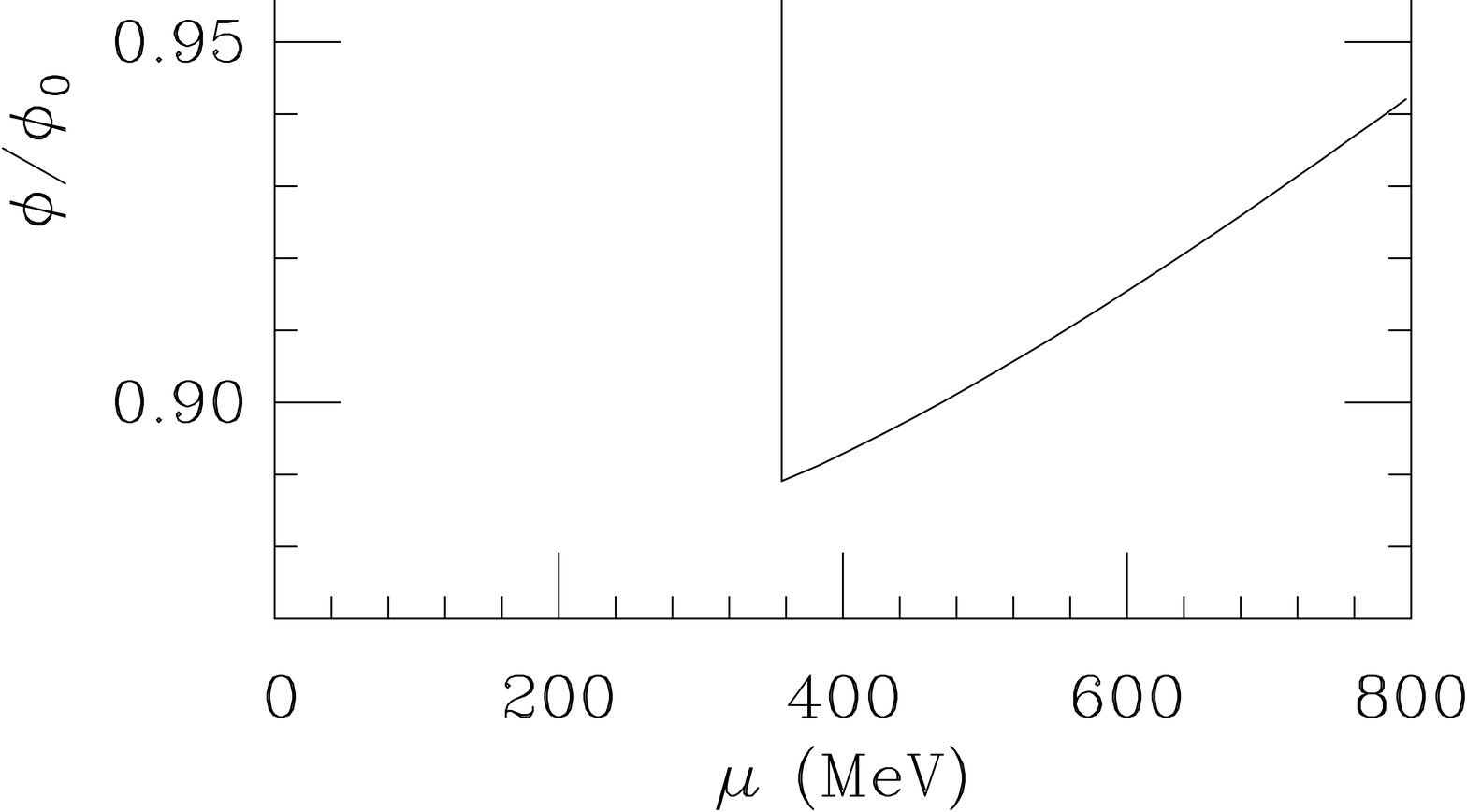,height=6cm}}
\centerline{Fig.~6}
\end{figure}

\end{document}